\documentclass[12pt]{article}
\usepackage{epsfig}
\usepackage{amsfonts}
\usepackage{amscd}
\usepackage{latexsym}
\usepackage{amsmath,amssymb}
\usepackage{verbatim}
\usepackage{color,soul}
\usepackage{color}
\usepackage{cite}
\usepackage[textheight=9in, textwidth=6.5in, letterpaper]{geometry}

\usepackage{hyperref}
\hypersetup{
    colorlinks,
    citecolor=black,
    filecolor=black,
    linkcolor=black,
    urlcolor=black
}

\usepackage{graphicx}
\usepackage{tikz}
\usepackage{tikz-3dplot}
\usetikzlibrary{calc}
\usetikzlibrary{positioning}
\usepackage{float}
\usepackage{subfig}
\usepackage[T1]{fontenc}
\usepackage{amsmath, amssymb, amsthm, fullpage}
\usepackage{doc}
\usepackage{ifthen, color, fancyvrb}
\usepackage{nextpage}

\usepackage{wrapfig}
\usepackage{parskip}
\usepackage{authblk}
\usepackage[makeroom]{cancel}

\numberwithin{equation}{section}
 
 \def\p{\partial}
 \def\bz{{\bar z}} 
   
  \def\ip{${\cal I}^+$}
\def\0{{(0)}}
\def\1{{(1)}}
\def\2{{(2)}} 

\def\cc{{\mathcal C}}

\def\ci{{\mathcal I}}

\def\cp{{\mathcal P}} 
\def\cw{{\mathcal W}}

\def\<{\langle }
\def\>{\rangle}

\newcommand{\bea}{\begin{eqnarray}}
\newcommand{\eea}{\end{eqnarray}}
\newcommand{\be}{\begin{equation}}
\newcommand{\ee}{\end{equation}}
\newcommand{\ba}{\begin{align}}
\newcommand{\ea}{\end{align}}
\def\be{\begin{equation}}
\def\ee{\end{equation}}
\def\beq{\be\begin{array}{c}}
\def\eeq{\end{array}\ee}

\def\vx{{\vec x}}

\def\be{\begin{equation}}
\def\ee{\end{equation}}
\def\bz{{\bar z}}
\def\p{\partial}
\def\ci{{\cal I}}
\def\ip{${\cal I}^+$}

\def\be{\begin{equation}}
\def\ee{\end{equation}}
\def\bea{\begin{eqnarray}}
\def\eea{\end{eqnarray}}
\def\<{\langle }
\def\>{\rangle}

\begin{document}
\begin{titlepage}
\unitlength = 1mm

\ \\

\vskip 3cm

\begin{center}

{ \LARGE {\textsc{Measuring Color Memory in a Color Glass Condensate at Electron-Ion Colliders}}}

\vspace{0.8cm}
Adam Ball${}^1$, Monica Pate${}^1$, Ana-Maria Raclariu${}^1$, Andrew Strominger${}^1$ and Raju Venugopalan${}^2$

\vspace{1cm}

{\it ${}^1$ Center for the Fundamental Laws of Nature, Harvard University,\\
Cambridge, MA 02138, USA}

{\it ${}^2$ Physics Department, Brookhaven National Laboratory, Bldg. 510A,\\
 Upton, NY 11973, USA}

\vspace{0.8cm}

\begin{abstract}
  The color memory effect is the non-abelian gauge theory analog of 
  the gravitational memory effect, in which the passage of color radiation induces a net relative SU(3) color rotation of a pair of nearby quarks. It is proposed that this effect can be measured in the Regge limit of deeply inelastic scattering at electron-ion colliders. 

   \end{abstract}

\vspace{1.0cm}

\end{center}

\end{titlepage}

\pagestyle{empty}
\pagestyle{plain}

\def\vx{{\vec x}}
\def\p{\partial}
\def\po{$\cal P_O$}

\pagenumbering{arabic}

\tableofcontents

\section{Introduction}

The passage of a gravitational wave causes the separation distance between a pair of inertial test bodies to oscillate. 
After the wave has fully passed, the oscillations cease but the final separation between the two bodies differs from the initial one.  This is the gravitational memory effect \cite{Zeldovich1974,BraginskiiThorne1987,Christodoulou:1991cr}.  The universal formula for the DC separation shift has recently been shown \cite{Strominger:2014pwa} to be fully equivalent to Weinberg's soft graviton theorem \cite{Weinberg:1965nx}, and also to be a direct consequence of  the infinite-dimensional asymptotic BMS symmetry \cite{Bondi:1962px,Sachs:1962wk} of general relativity. As such the memory effect measures subtle and rich features of general relativity in the deep infrared.

The oscillations of test bodies caused by gravitational waves have been definitively measured by LIGO\cite{Abbott:2016blz}. It is hoped that the memory effect will soon be measured \cite{Lasky:2016knh}. An obstacle to measuring the memory effect is that the LIGO test-body-mirrors are on pendula which eventually swing them back to their initial position, obscuring the DC memory effect.  Observation is nevertheless expected to be possible because of a separation of scales: the timescale of the pendula is longer than that of typical gravity wave signals, albeit not by terribly much.\footnote{The timescale of the LIGO pendula is about 100 msec, while most of the power output (which sources the memory) from GW150914 occurs over the slightly shorter scale 25 msec. eLISA  should be excellent for measuring memory.} 

Fully analogous memory effects also occur in QED and Yang-Mills theory,  where they enjoy the same triangular equivalence to soft theorems and asymptotic symmetries. In classical Yang-Mills theory, a pulse of color radiation incident on  an initially color singlet  pair of nearby test quarks or `quark dipole' causes oscillations in the relative quark color.
After the radiation pulse passes, a permanent relative color rotation remains and the quark dipole is no longer a color singlet. This is the `color memory' effect.  In exact analogy to the gravity case, the universal formula for the color rotation  has been shown \cite{Pate:2017vwa} to be fully equivalent to the soft gluon theorem  \cite{Berends:1988zn}, and also to be a direct consequence of recently discovered asymptotic Kac-Moody  symmetries  \cite{He:2015zea} of classical Yang-Mills theory.

One might expect that in QCD, which is based on SU(3) Yang-Mills theory, confinement would obstruct any measurement of the color memory effect. However, confinement may be viewed as an analog of the restoring pendula at LIGO: one must  find a situation in which a separation of scales allows the color rotation to be measured before confinement sets in. 
In this paper we will argue that high-energy   scattering  of heavy ions and electrons  provides such a venue, in which the separation of scales arises when the Lorentz-contracted ion radius is small compared to the QCD length. 

We consider deeply inelastic electron--heavy ion scattering in the Regge limit, where the momentum exchange is small compared to the center-of-mass energy. The physical picture of the process begins with the electron emitting  a virtual photon which splits into a quark-antiquark pair forming  a color singlet dipole. The quark dipole then traverses the ion and acquires net color. There is considerable experimental and theoretical evidence that this portion of the scattering process is well-approximated  by the Color Glass Condensate (CGC) effective field theory \cite{Iancu:2003xm,Gelis:2010nm,Kovchegov:2012mbw,Blaizot:2016qgz}, in which the region around the ion is described by classical Yang-Mills theory (corrected by running of the coupling)  sourced by the hard partons inside the ion. In this picture the quark dipole plays the role of the test quarks and crosses the highly relativistic Lorentz-contracted ion in a retarded time short compared to the QCD scale.  The change in the quark dipole color charge as it crosses the CGC region surrounding the ion is fully determined by the color memory effect. The cross-section for this stage in the scattering process is determined by the ``survival probability'' for the dipole to remain in a color singlet. Hence it is determined by the color memory effect. 

As discussed herein, collider data to date are encouraging but do not decisively confirm this picture  \cite{Rezaeian:2012ji,Kowalski:2007rw,Marquet:2007vb,Adare:2011sc,Lappi:2012nh,Dumitru,Dumitru:2010iy,Khachatryan:2010gv,CMS:2012qk,Chatrchyan:2013nka,Dusling:2013qoz,Adare:2015ctn,Dusling:2015gta,Aschenauer:2017jsk,Mantysaari:2017slo}.  Prospects for decisive confirmation of the color memory effect at a proposed future Electron-Ion Collider are excellent.

The paper is organized as follows.  In section 2, we review relevant aspects of the classical color memory effect presented in \cite{Pate:2017vwa}. We
start by reviewing the differential equation for the gauge transformation relating physically inequivalent vacua and then recall how this gauge transformation determines the classical relative color rotation of colored `test' quarks. In section 3, we describe how a classical color memory effect arises in the CGC effective field theory. We show that the same differential equation for a gauge transformation naturally emerges in the CGC framework and clarify
why it may be regarded as  a vacuum transition.  In section 4, we consider scattering processes in which the CGC theory is expected to apply and explain how the associated observables depend on
the color memory effect.  The extent to which the effect may have already been measured and the prospects for definitive future confirmation are discussed. 

\section{Color memory review}
\label{cmr}
In this section we review the color memory effect in the form derived in \cite{Pate:2017vwa}. We work in retarded coordinates $(u,r,z,\bz)$ in which the Minkowski metric reads
\bea
\label{rc}
ds^2&=&-dt^2+(dx^1)^2+(dx^2)^2 +(dx^3)^2\cr&=&  -du^2 - 2du dr + 2 r^2 \gamma_{z\bz} dz d \bz, \quad \quad \quad \gamma_{z \bz} = \frac{2}{(1+z \bz)^2}.
\eea
 The retarded coordinates are related to the Cartesian coordinates $(t, x^1, x^2, x^3)$ by
\be
	\begin{split}
	t = u + r, \quad \quad \quad & (x^1, x^2, x^3) 
	= r\left(\frac{z+ \bz}{1+ z \bz}, \frac{-i(z -\bz)}{1+ z \bz}, \frac{1-z \bz}{1+ z \bz}\right), \\  & (x^1)^2 + (x^2)^2 + (x^3)^2 = r^2.
	\end{split}
\ee
Future null infinity, denoted $\mathcal{I}^+$, is approached by taking the limit of $r \rightarrow \infty$ for fixed $u$.  In this limit, 
$(z, \bz)$ become stereographic coordinates on the `celestial sphere' at $\mathcal{I}^+$.

Consider a pair of classical (unconfined) test `quarks' at $\ci^+$ subject to an $SU(3)$  Yang-Mills field with non-abelian field strength $F_{\mu\nu} = F_{\mu\nu}^a T^a$ and gauge potential $A_{\mu} = A_{\mu}^a T^a$, where $T^a$ are the generators of the gauge group $SU(3)$ in the triplet  representation. The color of a quark $q$ moving along a trajectory with tangent vector $u^\mu$ evolves according to  
\begin{equation}
\label{pt}
u^{\mu}(\nabla_{\mu}q  - ig_{\rm YM} A_{\mu} q) = 0,
\end{equation}
where $g_{\rm YM}$ is the coupling. 
$A_{\mu}$ obeys the Yang-Mills equation sourced by a matter current $j_{\mu}^M$
\begin{equation}
\label{YMeq}
\nabla^{\nu}F_{\nu\mu} - i g_{\rm YM}[A^{\nu}, F_{\nu\mu}] =  j_{\mu}^{M}.
\end{equation}
Under the passage of a finite-duration color radiation flux through $\ci^+$, the pair of quarks will acquire a relative color rotation given by the solution to \eqref{pt}. To describe  this explicitly, it is convenient to work in temporal gauge, $A_u = 0$, and assume that before some initial retarded time $u_i$, the pair of quarks is in the vacuum characterized by the  flat connection 
$\left. A_z(u_i,r,z,\bz)\right|_{\ci^+} = 0$.  During a finite interval $u_i< u<u_f$, color flux passes through $\ci^+$, after which the system returns to vacuum and the gauge connection  on the celestial sphere is again flat, but not vanishing.
Given our  $A_z = 0$ starting configuration, the change in the gauge connection over an interval $u_i< u<u_f$ is just the flat connection at the final time $\left. A_z(u_f,r,z,\bz)\right|_{\ci^+} =\frac{1}{g_{\rm YM}} i \, U\p_z U^{\dagger}|_{u_f}$. 
Solving \eqref{YMeq} in a large-$r$ expansion, this is given by the solution to the non-linear differential equation on the sphere
\begin{equation}
\label{de}
-  \Delta \left(D_{\bz}  A_z^{(0)} + D_{z}A_{\bz}^{(0)}\right)  = \gamma_{z\bz}\left(\int_{u_i}^{u_f}du J^{(2)}_u + \Delta F_{ru}^{(2)}\right),
\end{equation}
where $\Delta A$ denotes the change in flat connections between $u_i$ and $u_f$. 
Here
\begin{equation}\label{rsx}
J _u = ig_{\rm YM}\frac{\gamma^{z\bz}}{r^2}\left([A_z ,F_{\bz u} ] + [A_{\bz} , F_{zu} ] \right) +  j_u^{M }, 
\end{equation} 
for any field $\mathcal{O}(r,u,z,\bz)$ we define $\mathcal{O}^{(n)}(u,z,\bz)$ 
as the coefficient of the $r^{-n}$ term in its large-$r$ expansion about $\mathcal{I}^+$, and $D_z$ is the covariant derivative with respect to $\gamma_{z \bz}$.
 \eqref{de} was solved  perturbatively in a weak-field expansion in \cite{Pate:2017vwa}.

Consider a pair of quarks, initially coincident at point $(z_1, \bz_1)$ on the celestial sphere in a color singlet. Then at or before    $u = u_i$ we separate one of the quarks to $(z_2, \bz_2)$, which involves no color rotation because we are taking the initial connection to vanish. They then evolve with no color rotation (because $A_u=0$) from  $u_i$ to $u_f$ along $\ci^+$ 
fixed at their respective points on the sphere
and are recombined at time $u_f$ back at $(z_1, \bz_1)$. Since according to  \eqref{de} the final connection is flat but nonvanishing,  the recombined quarks are no longer in a color singlet. Rather they acquire the relative color rotation 
\be \label{colorrot}
	U(z_1, \bz_1) U^\dagger(z_2, \bz_2) 
=
U (z_1, \bz_1; z_2, \bz_2) = \mathcal{P}\exp \left(i g_{\rm YM}\int_{(u_f, z_2,\bz_2)}^{(u_f, z_1,\bz_1)}A \right) ,
\end{equation}
where  $\mathcal{P}$ denotes path ordering, and there is no dependence on the path taken between the endpoints. This is the color memory effect. The relative color rotation of the recombined quarks measures the change in the flat gauge connection between $u_i$ and $u_f$, which can be interpreted as a transition between the vacua at $u_i$ and $u_f$ \cite{Strominger:2013lka,He:2015zea}. The formula is easily generalized (by conjugation) to the case of a flat but nonzero initial connection. 

This  analysis pertains directly to an  initially color singlet  `quark dipole', where $(z_1, \bz_1)$ and $(z_2, \bz_2)$ would be the locations of the two ends of the dipole and \eqref{colorrot} measures the amount by which the quark dipole fails to remain in a color singlet 
configuration. 

While the expression for this phase was obtained in a particular gauge in which the initial connection was flat and $A_u=0$,  we readily obtain a gauge covariant expression for the net color rotation,
whose trace is given by
\be
\label{wl} \cw_\cc  =\frac{1}{N_c} {\rm Tr}~ \cp \exp \left(i  g_{\rm YM}\oint_{ \cc} A\right)  ,
\ee
where ${\rm Tr} (T^a T^b) = \delta^{ab}$, $N_c =3$, and 
 $\cc$ is the closed contour on $\ci^+$ circling around the loop defined by the separated quark worldlines.
 
\section{Color memory in the CGC}
\label{cmcgc}

Remarkably, the conditions necessary for the color memory effect arise naturally in the description of high-energy scattering in the Regge limit of QCD.  
The Regge limit, which entails taking the center-of-mass energy $\sqrt{s}\rightarrow \infty$ at 
fixed (large) squared momentum transfer  $-t = Q^2\gg \Lambda_{\rm QCD}^2$, is both accessible and routinely achieved at colliders.  Rephrasing the Regge limit in deeply inelastic scattering as taking Bjorken $x_{\rm Bj}\sim Q^2/s\rightarrow 0$ at fixed  large $Q^2$
elucidates the physical regime probed in this limit as that of configurations in hadrons containing large numbers of partons, primarily gluons, each carrying a small fraction $x$ ($\approx x_{\rm Bj}$) of the hadron's momentum.\footnote{In deeply inelastic scattering, a lepton probes a hadron of momentum $P$ by exchanging a photon of momentum $q$. Bjorken $x_{\rm Bj}$ is defined by $x_{\rm Bj} \equiv -\frac{q^2}{2 P \cdot q}$, where $q^2 = Q^2$ is the squared four-momentum transferred.   $x_{\rm Bj}$ is  fixed by
 kinematics to coincide with the longitudinal momentum fraction $x= {p^+\over  P^+}$ carried by the struck parton.}
 
  Broadly speaking, the physical picture that emerges resembles that of the color memory effect. The relevant degrees of freedom are the `soft' small-$x$ gluons (or wee gluons) radiated by faster, localized partons at larger $x$.  Moreover,  since $Q^2 \gg \Lambda^2_{\rm QCD}$, the dynamics 
 of the wee gluons is weakly coupled and deconfined, albeit strongly correlated, because gluon occupancies  are  non-perturbatively large ($\sim 1/g_{\rm YM}^2$).  These considerations therefore suggest that our discussion in the previous section may be directly applicable to the physics of high energy hadron scattering in QCD. 
Indeed, we will show in this section that the color memory effect is contained in a classical effective field theory of QCD at small $x$, the Color Glass Condensate (CGC)~\cite{Iancu:2003xm,Gelis:2010nm,Kovchegov:2012mbw,Blaizot:2016qgz}. 

The CGC is formulated in the infinite momentum frame (IMF) $P^+\rightarrow \infty$ for a hadron moving in the $x^+$ direction where
	\be
		x^\pm = \frac{t \pm x^3}{\sqrt{2}}\,, ~~~~\vec x= (x^1,x^2)\
	\ee
	are the lightcone coordinates, and the flat metric \eqref{rc}  becomes 
	\be
	\label{lc}
		ds^2 = -2 dx^+ dx^- + d x^i dx^i, \quad \quad \quad i = 1, 2,
	\ee 
where the sum over the repeated transverse index $i$ is here and hereafter implied. 	These coordinates are related to the spherical retarded coordinates in \eqref{rc} by   \be
		\begin{split} \label{coordrelate}
			x^+ = \frac{1}{\sqrt{2}}\left(u + \frac{2 r}{1+ z \bz}\right), \quad \quad  x^- = \frac{1}{\sqrt{2}}\left(u + \frac{2r z \bz}{1+ z\bz}\right), \quad \quad   x^1+i x^2  = \frac{2rz}{1+z \bz}.
		\end{split}
	\ee
	The IMF is reached by taking $\lambda \to \infty$ with $(x^+,x^-) \to (\lambda x^+,\lambda^{-1} x^-)$ or equivalently  $(r,u,z,\bz) \to (\lambda r,\lambda^{-1}u,\lambda^{-1}z,\lambda^{-1}\bz)$. In this infinitely boosted frame  \eqref{coordrelate} reduces to
	\be \label{fct}
		x^+ =\sqrt{2}  r  , \quad \quad \quad x^- = \frac{1}{\sqrt{2}}\left(u + 2r z \bz \right), \quad \quad \quad x^1+i x^2  = 2rz ,
	\ee
	the celestial sphere at null infinity is flattened to a 2d plane, and the retarded metric reads
	  \be
	  \label{frc}
 	ds^2= -2dudr+4r^2dzd\bz.
 \ee
	Since the IMF is used to describe a hadron moving in the $x^+$ direction, evolution in $x^+$ is naturally interpreted as evolution in time.  
		The utility of this frame  is that it clearly exhibits the natural separation of scales appearing in the Regge limit.  In particular, a parton with lightcone momentum $k^+ = x P^+$ has typical $x^+$  lifetime 
		\be
			 \Delta x^+ \sim \frac{1}{k^-}    = \frac{2k^+}{{m}_\perp^2}  =x \frac{2P^+}{{m}_\perp^2} , 
		 \ee
		 where $m_\perp^2 \equiv k_ik_i + m^2$ and longitudinal spread
		 \be \label{ls3}
		 	\Delta x^- \sim \frac{1}{k^+}   = \frac{1}{x P^+}.
		 \ee
		 As such, large-$x$ (or `hard') modes with lightcone momenta $k^+\sim P^+$  are vastly time dilated relative to the small-$x$ (or `soft') modes with longitudinal momenta $k^+ \ll P^+$.  For the purposes of small-$x$ dynamics, the problem can be formulated 
		 as an effective field theory~\cite{McLerran:1993ni,McLerran:1993ka,McLerran:1994vd}--the CGC--wherein large-$x$ modes are treated as static color sources coupled to the  classical gauge fields at small $x$.
Likewise, the large-$x$ color sources are highly Lorentz contracted in the $x^-$ direction~\cite{Bjorken:1982qr,McLerran:1993ni,Iancu:2003xm} with support $\sim 1/P^+$.
 Working in the gauge
\be
		A_+ = 0 ,
	\ee
the static color current can therefore be approximated by a shockwave travelling in the $x^+$ direction: 
	\be \label{jmu}
		j^\mu = \delta^{\mu +} \delta (x^-) \rho(\vec{x} ), \quad \quad \quad  \rho  = \rho^a T^a.
	\ee
To first approximation, we can ignore time-dependent fluctuations of the wee gluons, so that  their configurations are just given by static ($x^+$-independent)
classical solutions to the Yang-Mills equation with color current source \eqref{jmu}. 
We will return to this point later in this section,
but the upshot is that the full power of the CGC as an effective theory lies in the fact that it justifies this approximation over a range in parameter space.  

  Working in $A_+ = 0$ gauge and taking the gauge fields to be $x^+$-independent, the Yang-Mills equations simplify to
\be
	\label{YMe}
			\begin{split}
				 \p_i F_{i-} - ig_{\rm YM}[A_i, F_{i-}] &=   -\delta (x^-) \rho (\vec{x})\,,\\  
				 \p_j F_{ji} -ig_{\rm YM}[A_j, F_{ji}]&=0\,.
			\end{split}
	\ee
	 The second equation in \eqref{YMe} can be immediately solved by setting $F_{ij} = 0$, $i.e.$ no magnetic fields. This turns out to be the physically relevant case and we restrict to such field configurations from now on.

Further taking  $A_-=0$, the first equation  in \eqref{YMe} becomes 
 \begin{equation}
 \label{YMtg}
 \p_-\p_i A_i -  i g_{\rm YM} [A_i,\p_-A_i] =  \delta (x^-) {  \rho} (\vec{x})\,.
 \end{equation}
In terms of the net color current  \eqref{rsx} $J_- = -\delta(x^-){  \rho} (\vec{x}) - i g_{\rm YM} [A_i,\p_-A_i]$,   \eqref{YMtg} integrates to 
\begin{equation}
\label{del}
-\p_i\Delta A_i = \int_{x^-_i}^{x^-_f} dx^- J_-\,,
\end{equation}
where we assume that  $x^-_i<0<x_f^-$  and $\Delta A_i$ is the change
in flat connections on the transverse plane between $x^- = x^-_i$ and $x^- = x^-_f$.  

\eqref{del} is 
clearly  a special case of the central memory equation \eqref{de}.  
First we note that since \eqref{del} is $x^+$-independent we may take it to \ip\ simply by shifting $x^+$. 
Second, since \eqref{del} was derived in the IMF, it must be compared to a memory equation  in an infinite momentum frame. As discussed above, such a frame can be reached by rescaling the retarded radial coordinates so
that the celestial sphere is flattened to a plane  with $\gamma_{z \bz} = 2$ as in \eqref{frc}.  
The above-imposed static condition in $A_+=0$ gauge implies we are solving the equation with $\Delta F_{ru}^{(2)} = 0$.  Setting $\Delta F_{ru}^{(2)}$ to zero in  \eqref{de}, it transforms into  \eqref{del} under the coordinate transformation \eqref{fct} from  \eqref{frc} to  \eqref{lc}. This establishes the equality  of the large gauge transformations $\Delta A$ encountered in CGC with those of the memory effect.  
	 
	 Thus far, by working in the IMF, we have shown how deconfined gluons in vacuum-to-vacuum field configurations appear in the Regge limit of QCD.  However, these arguments implicitly rely on a weak-coupling description
	 in which quantum fluctuations of wee gluons are parametrically suppressed.  Nevertheless, the full effective field theory framework that comprises the CGC can be used to justify these assumptions.
	 To begin with, the separation of partons into large-$x$ sources and small-$x$ gluons inevitably introduces a cutoff $\Lambda^+$ that distinguishes the two.  The requirement that physical observables do not depend on the ``arbitrary''  separation between large $k^+> \Lambda^+$ and small $k^+< \Lambda^+$ modes lends itself to a Wilsonian renormalization group (RG) treatment ensuring independence of observables on $\Lambda^+$: small quantum fluctuations around the classical solutions ($g_{\rm YM}^2\log(\Lambda^+/{\Lambda^\prime}^+)\ll 1$) can be integrated out generating the same classical theory at scale $\Lambda^\prime$ with $\rho\rightarrow\rho^\prime$~\cite{JalilianMarian:1996xn}.
Importantly, the classical theory at the new scale $\Lambda'$ still exhibits vacuum transitions because although $\rho'$ has increased support in the $x^-$ direction   compared to $\rho$, the   classical  wee gluons at the new scale are 
	 guaranteed to  be delocalized over an even larger distance in $x^-$ by virtue of the new cutoff $\Lambda'$.  
	Hence  renormalization group evolution can be viewed as a function of  $x$.
	 
	  We now turn to some specific solutions of \eqref{YMe}.
	The solutions to \eqref{YMtg} of interest in the CGC are those obtained by regularizing the delta function in $x^-$ (understood to have support $\sim 1/\Lambda^+$).
	Without loss of generality, we take the source to have support over the range $1/P^+< x^-< 1/\Lambda^+$, where $P^+ \gg  \Lambda^+$.
	 We introduce a new coordinate (defined for $x^-\geq0$), the spacetime rapidity  
	\be
	\label{rapidity-def}
	\eta=\eta_{\rm source} - \log(x^-/x_{\rm source}^-),
	\ee
	where
	\be
		\eta_{\rm source} \equiv \frac{1}{2} \log \frac{\Lambda^+}{\Lambda^-}, \quad \quad \quad x^-_{\rm source} \equiv \frac{1}{\Lambda^+}, \quad \quad \quad \Lambda^- \equiv \frac{\Lambda_\perp^2}{2 \Lambda^+},
	\ee
	and $\Lambda^2_\perp$ is a typical transverse momentum scale.  One then  finds
	\be
		\eta \big|_{x^- = \frac{1}{x P^+}} = \eta_{\rm hadron}  + \log x, \quad \quad \quad \eta_{\rm hadron} \equiv \frac{1}{2} \log \frac{P^+}{P^-} =\log  \frac{\sqrt{2}P^+}{\Lambda_\perp}.
	\ee
	The sources in the two coordinate systems are related by 
	 \be
	  	x^- \rho(x^-, \vec{x}) = - \rho_{\eta}(\eta, \vec{x}).
	  \ee 
	$F_{ij}=0$ implies that the transverse component of the gauge potential takes the form 
	\begin{equation}
A_i(\eta , \vec{x}) =\frac{ i}{g_{\rm YM}}U(\eta ,\vec{x})\p_i U^{\dagger}(\eta,\vec{x})\,.
\end{equation}
$A_i$ may be set to zero by a residual gauge transformation which preserves our gauge condition $A_+=0$ but generates a nonzero $A_\eta$. Doing so leads to a useful  formula for $A_i$  in terms of the sources in the $A_i=0$ gauge, as follows \cite{JalilianMarian:1996xn}. 
 When  $A_i = 0$, the first equation in \eqref{YMe}  
reduces to the Poisson equation in the transverse plane and can be solved for $A_\eta$ via standard Green's function methods~\cite{JalilianMarian:1996xn,Kovchegov:1996ty}:
 \bea
 \label{pe}
 A_\eta(\eta,\vec{x}) &=&\int d^2\vec {y} G(\vec{x}-\vec{y}){\tilde \rho}_{\eta}(\eta, \vec{y}),\cr \qquad G(\vec{x}-\vec{y}) = -\frac{1}{2\pi}\log|\vec{x} - \vec{y}|, &&\qquad \vec \p^2  G(\vec{x}-\vec{y}) = -\delta^{(2)}(\vec{x} - \vec{y}), \eea
 where $\tilde \rho_{\eta}$ differs from $\rho_{\eta}$ appearing previously by the gauge transformation used to set $A_i = 0$ (see \eqref{newYM} that follows).   $A_\eta$ may be set back to 0 by a gauge transformation
\be
	A_\mu \to  g \left (A_\mu + \frac{i}{g_{\rm YM}}  \p_\mu \right )g^\dagger,
\ee
where
\be \label{eikonalphase}
	g(\eta, \vec{x}) = 
		\mathcal{P}\,\exp\left(i g_{\rm YM}\int^{\eta_i}_{\eta} d{\eta'} \int d^2\vec{y} G(\vec{x}-\vec{y}) \tilde  \rho_{\eta}({ \eta'},\vec{y}) \right)
		\equiv U_{\eta_i,\eta}(\vec{x}).
\ee
One thereby  obtains an  expression for $A_i$ in terms of the data $\tilde \rho_{\eta}$
\begin{equation}
\label{WW-field}
A_i (\eta,\vec{x}) = \int d^2 \vec{y} \, \partial_i G(\vec{x}-\vec{y}) \int^{\eta_i}_{\eta} d\eta^\prime 
U_{\eta_i,\eta'}(\vec{x}) \tilde \rho_{\eta}(\eta^\prime,\vec{y}) U_{\eta', \eta_i}(\vec{x}) \,. 
\end{equation} 
For completeness, notice that the Yang-Mills equation solved by this expression is
\be \label{newYM}
	\mathcal{D}_i \frac{d A_i}{d\eta} =\rho_{\eta} (\eta,\vec{x}) = U_{\eta_i,\eta }(\vec{x}) \tilde \rho_{\eta}(\eta ,\vec{x}) U_{\eta ,\eta_i}(\vec{x}),
\ee
where $\mathcal{D}_i = \p_i - i g_{\rm  YM} [A_i, \cdot]$ is the gauge-covariant derivative. 

In \eqref{WW-field}, the gauge field at the rapidity $\eta$ depends on the sources at larger rapidities; this equation can therefore be solved iteratively from an initial condition set at a rapidity $\eta_i$ beyond which the source density   
has no
support (for example $\eta > \eta_{\rm hadron}$). As such, $\eta_i$ will correspond to the initial retarded time $u_i$ from section 2 where the transverse component of the gauge field vanishes. Likewise, the source vanishes at rapidities $\eta< \eta_{\rm source}$, and the gluon fields at these rapidities describe the final vacuum configuration which is generically different from the initial
vacuum set by the initial condition at large rapidity. 

$U_{\eta_i, \eta}(\vec{x})$ defined in \eqref{eikonalphase} is simply related to the color rotation \eqref{colorrot}. To see this, let us determine the relative color rotation when $A_u$ or equivalently $A_{\eta}$ is the only non-zero component. In this case, the contributions to the relative color rotation of a quark dipole come from the segments of the worldlines extending in the $\eta$ direction at fixed separated points in the transverse plane. This color rotation is related to  \eqref{colorrot} by a gauge transformation that sets $A_{\eta} = 0$
\begin{equation}
g (\eta_f,\vec{x}_1 ) U^{\dagger}_{\eta_i,\eta_f}(\vec{x}_2)U_{\eta_i,\eta_f}(\vec{x}_1)g^{\dagger}(\eta_f,\vec{x}_1) = U(z_1,\bz_1)U^{\dagger}(z_2,\bz_2),
\end{equation}
where $\vec x$ is related to $(z, \bz)$ by the coordinate transformation \eqref{fct} and $\eta$ is related to $u$ by \eqref{fct} and \eqref{rapidity-def}.
Simplifying this using \eqref{eikonalphase} we find this becomes 
\begin{equation}
\label{compare-rot}
U_{\eta_i,\eta_f}(\vec{x}_1)U^{\dagger}_{\eta_i,\eta_f}(\vec{x}_2) = U(z_1,\bz_1)U^{\dagger}(z_2,\bz_2).
\end{equation}

The prescription for computing physical observables in CGC is to perform an average over color charge configurations,
\begin{equation}
\langle {\cal O}\rangle = \int [D\rho] \,W_{\Lambda^+}[\rho]\, {\cal O}[\rho] \,,
\label{weight}
\end{equation}
where the weight functional $W_{\Lambda^+}[\rho]$ represents the distribution of the large-$x$ color charge densities. If one assumes that the sources are uncorrelated, as in a large nucleus for instance~\cite{McLerran:1993ni,Kovchegov:1996ty,Jeon:2004rk}, then the weight functional can be taken to be a Gaussian distribution 
\begin{equation}
W_{\Lambda^+}[\rho_{\eta}] = N \exp\left(-\int_0^\infty d\eta \int d^2 \vec{x} \frac{{\rm Tr} \rho_{\eta}(\eta,\vec{x})^2}{2 \mu_{\Lambda}^2(\eta,\vec{x})}\right), 
\label{Gaussian}
\end{equation}
where $N$ is a normalization constant. 
Here $\mu^2_{\Lambda}(\eta,\vec{x})$, the color charge squared per unit rapidity per unit area, is a dimensionful scale in the CGC EFT. As we shall discuss further in the next section, it can be fixed by computing a physical observable, and thence used to compute other observables. 
As mentioned previously, the requirement that $\langle {\cal O}\rangle$ be independent of $\Lambda^+$ generates a Wilsonian RG equation in $x$ for $W_{\Lambda^+}$, the JIMWLK equation~\cite{JalilianMarian:1997gr,JalilianMarian:1997dw,Iancu:2000hn,Ferreiro:2001qy}. In a limit of low parton densities $\rho/|\vec{k}|^2 \ll 1$, the JIMWLK equation simply reduces to the BFKL equation~\cite{Kuraev:1977fs,Balitsky:1978ic} that resums all the radiative emission and virtual Feynman graphs of perturbative QCD in the leading logarithmic $g^2_{\rm YM}
\log(1/x)\sim 1$ approximation. As a result of the RG procedure, 
the structure of the classical equations we described here is unchanged by these quantum corrections. 
Namely,    quantum fluctuations of gluons, which become large at small $x$ and would in principle spoil the classical approximation, are recast as statistical fluctuations of the classical source $\rho$. 
 Hence,
their primary effect~\cite{Dumitru:2011vk} is   
the evolution of $\mu^2_\Lambda$ with  respect to $x$,  which is described by a nontrivial closed form nonlinear evolution equation--the Balitsky-Kovchegov equation~\cite{Balitsky:1995ub,Kovchegov:1999yj}. 

In any case,  the weight function $W_{\Lambda^+}[\rho]$ introduces a dimensionful scale in the CGC EFT, which ultimately appears in CGC observables.   
In the case of a Gaussian form of the weight functional as in \eqref{Gaussian}, the scale set by $\mu_\Lambda$ is a measure of the size of the color fluctuations of the source $\rho_{\eta}$ per unit area, per unit rapidity.  Hence, 
 it is in turn directly related to the size of the color memory effect.  This relation will be explained in more detail in the following section. 

The above discussion is key to measuring the color memory effect in QCD since we are unable to directly observe colors of particles.  
In the next section, we will see that the QCD analog of the probes, which serve as the analog of the pair of nearby inertial detectors in the gravitational memory effect, are color dipoles that are sensitive to the color flux from the target long after the interaction and can be simply related to physical observables at collider energies. 

\section{Measuring color memory in deeply inelastic scattering}

In general, a pair of quarks interacting with the shockwave field will be color rotated according to \eqref{colorrot}. For the specific case of the  CGC, the pair will be color rotated by \eqref{compare-rot}. 
This rotation is not directly observable in QCD. However cross-sections in Regge asymptotics are sensitive to the color averaged product of the rotation at a given transverse spatial location $\vec{x}$ and its Hermitian conjugate at a different spatial location $\vec{y}$. 

The simplest and cleanest measurement of color memory is in deeply inelastic scattering (DIS) off protons and nuclei. At high energies, the spacetime picture is that of the virtual photon emitted by the electron splitting into a quark-antiquark pair that subsequently scatters coherently off the shockwave~\cite{Gribov:1965hf,Ioffe:1969kf,Bjorken:1970ah,McLerran:1998nk,Balitsky:2001mr,Roy:2018jxq}. 
In the eikonal approximation, the effect of the shockwave is to induce a color rotation on a quark at $\vec{x}$ by $U(\vec{x})\equiv U_{-\infty,\infty}(\vec{x})$ in \eqref{eikonalphase}. Combining this with the corresponding formula for an antiquark at $\vec{y}$ and using \eqref{compare-rot} we obtain the scattering amplitude for a color dipole
\begin{equation}
\label{sa}
\mathcal{S}(\vec{x},\vec{y}) = \frac{1}{N_c}\text{Tr}\left(U(\vec{x})U^{\dagger}(\vec{y}) \right)=  \cw_\cc,
\end{equation}
 where     the amplitude 
 is normalized to 
$1$ in the case where $U$ is the identity matrix in the fundamental representation. 
Here we see that the the color memory effect -- or quark dipole color misalignment -- is directly responsible for the leading term in the dipole scattering amplitude! 
Since this amplitude pertains to  the forward scattering limit, it 
is simply related to the dipole cross-section by the optical theorem:
\begin{equation}
\sigma_{\rm dipole}(x,\vec{r}) = 2\int d^2 \vec{b} \left[1- \langle  {\rm Re}\ \mathcal{S}(\vec{x},\vec{y})\rangle\right],
\label{dipole}
\end{equation}
where $\vec{r}=\vec{x}-\vec{y}$, $\vec{b}=(\vec{x}+\vec{y})/2$, $x\approx x_{\rm Bj}$ (up to higher order corrections) and $\langle \cdots\rangle$ corresponds to the expectation value 
in \eqref{weight}. The 
$x$ dependence enters through the averaging over color charge configurations.

The inclusive DIS virtual photon-hadron cross-section $\sigma_{\gamma^* H}(x,Q^2)$ factorizes into a piece corresponding to the splitting of the virtual photon into a quark-antiquark pair and the dipole scattering cross-section.  
It can be expressed as
~\cite{McLerran:1998nk,Venugopalan:1999wu,Kovchegov:1999yj} (see also \cite{Nikolaev:1990ja,Mueller:1989st})  
\begin{equation}
\sigma_{\gamma^* H}(x,Q^2) = \int_0^1 dz\int d^2 \vec{r} |\Psi(z,\vec{r},Q^2)|_{\gamma^*\rightarrow q\bar{q}}^2\,\sigma_{\rm dipole}(x,\vec{r})\,,
\end{equation}
where   $\Psi(z,\vec{r},Q^2)_{\gamma^*\rightarrow q\bar{q}}$ is the lightcone wavefunction~\cite{Bjorken:1970ah} for a photon splitting into a $q\bar{q}$ pair of transverse size $|\vec{r}|$, with the quark carrying a fraction $z$ of the virtual photon longitudinal momentum. Thus while $\cw_\cc$  is not directly observable, we see however that the inclusive DIS cross-section in the high energy limit can be simply related to the color memory effect. 

 To understand the effect of the statistical averaging procedure, let us calculate \eqref{dipole}
assuming a Gaussian distribution \eqref{Gaussian} of color sources.  If the color sources are Gaussian distributed, then all correlation functions can be written in terms of
the two-point function
\begin{equation}
\label{stpf}
\langle \rho^a(x^-, \vec{x}) \rho^b(y^-,\vec{y})\rangle = \delta^{ab} \delta(x^- - y^-)\delta^{(2)}(\vec{x} - \vec{y})\mu^2 (x^-,\vec{b}).
\end{equation}
 It follows that the dipole scattering amplitude in \eqref{dipole} can be obtained by exponentiating the leading order term arising from an expansion  of \eqref{eikonalphase}. Taking $\mu^2$ to be $\vec{b}$-independent, the result is
\begin{equation}
\label{dsm}
\langle \mathcal{S}(\vec{r})\rangle = \exp\left\{-g_{\rm YM}^2 C_F \int \frac{d^2\vec{k}}{(2\pi)^2} \frac{1 - e^{i{\vec{k}}\cdot {\vec{r}}}}{{|\vec{k}|}^4}\int dx^- \mu^{2}(x^-)\right\} \approx \exp\left\{-\frac{1}{4} |\vec{r}|^2 Q_s^2 \right\},
\end{equation}
where $C_F$ is the quadratic Casimir in the fundamental representation and $Q_s$ is the saturation scale
\begin{equation}
\label{ss}
Q_s^2 = \frac{g_{\rm YM}^2}{4\pi}C_F \log\left(\frac{1}{\Lambda_{\rm QCD}^2 |\vec{r}|^2}\right)\int dx^- \mu^2(x^-).
\end{equation}
The QCD scale $\Lambda_{\rm QCD}$ arises as an infrared cutoff for the logarithmically divergent integral in \eqref{dsm} while the dipole separation acts as a UV cutoff $|\vec{k}| \sim 1/|\vec{r}|$. 

We see $Q_s$ is proportional to the color charge density $\mu$ of the sources and moreover, it weakly depends on the transverse resolution $|\vec{r}|$ of the probe. For fixed dipole size $|\vec{r}|$, $\sigma_{\rm dipole}$ increases with $\mu$, meaning that larger color charge densities of the sources result in a stronger deflection of the dipole and therefore indicates an enhanced color memory effect. 

For fixed $Q_s$, \eqref{dipole} and \eqref{ss} show that small size dipoles are less affected by the field distribution of the nucleus, while large dipoles are strongly absorbed. The transition between these behaviors occurs at $|\vec{r}| \sim 1/Q_s$. This scale coincides with the scale at which the nucleus becomes densely populated with strong gluon fields of typical momenta $\sim Q_s$. The inverse of the saturation scale also plays the role of the length scale at which color charges of effective gluon distributions are screened. Moreover, it is the typical transverse momentum of closely packed wee gluons in the hadron ~\cite{Gribov:1984tu,Mueller:1985wy,Mueller:2001fv}. This implies that the coupling should be evaluated at $Q_s$, and the assumption is that the CGC applies as an EFT when $Q_s$ is sufficiently large in comparison with $\Lambda_{\rm QCD}$.

A consequence of the nontrivial $A^i$ generated by the color memory effect is that the correlator $\langle A^i A^i\rangle$, computed using \eqref{weight} and \eqref{Gaussian}, is directly proportional to the non-abelian Weizs\"{a}cker-Williams distribution of gluons in the hadron~\cite{McLerran:1993ni,McLerran:1993ka,JalilianMarian:1996xn,Kovchegov:1996ty}. The corresponding color field strengths are characterized by a typical saturation momentum scale $Q_s\propto \mu$ where $\mu$ is the color charge density of sources per unit volume in \eqref{Gaussian}; a high energy probe experiencing an instantaneous interaction with the gluon shockwave will therefore receive a transverse momentum kick of order $|\Delta \vec{p}|\sim Q_s$, of which it will retain the color memory until the much longer timescales of hadronization. This is the non-abelian analog of the electromagnetic memory effect \cite{Bieri:2013hqa,Pasterski:2015zua}. 

In fact, it is precisely the color rotation \eqref{eikonalphase} that
 gives rise to a transverse momentum kick of a quark propagating through a target. This is because \eqref{sa} is also related to the amplitude for a quark (or antiquark) to acquire a momentum kick $\Delta \vec{p} $ when scattering off a nucleus. Summing over final states of the nucleus and quark colors as well as averaging over initial colors \cite{Blaizot:2016qgz}, 
the associated probability $\mathcal{P}(\vec{b}, \Delta \vec{p})$ is
\begin{equation}
\int d^2\vec{b} \mathcal{P}(\vec{b},\Delta \vec{p}) = \int d^2\vec{b} \int d^2\vec{r} e^{-i\Delta \vec{p}\cdot \vec{r}} \langle \mathcal{S}(\vec{r})\rangle.
\end{equation}
Using the dipole scattering amplitude found before, this probability reduces to  
\begin{equation}
\label{mkp}
\mathcal{P}(\vec{b}, \Delta \vec{p}) \approx \int d^2\vec{r} e^{-i\Delta \vec{p}\cdot \vec{r}} e^{-\frac{1}{4}|\vec{r}|^2 Q_s^2} \approx \frac{4\pi}{Q_s^2} e^{-|\Delta \vec{p}|^2/Q_s^2},
\end{equation}
where the dependence of $Q_s$ on $|\vec{r}|$ was ignored, a good approximation for large enough $|\vec{r}|$.\footnote{At small $|\vec{r}|$ or equivalently large transverse momenta, \eqref{mkp} has a power law decay with $|\Delta \vec{p}|$. }
 This is a Gaussian distribution centered at $\Delta \vec{p} = 0$ whose width is proportional to the saturation scale and whose amplitude decreases as $1/Q_s^2$. It follows that larger $Q_s$ lead to larger momentum kicks in agreement with our previous interpretation of $Q_s$ as associated with the strength of the color memory effect.

 While we have just shown that there is a simple interpretation of the two-point dipole correlator $\langle \text{Tr}\left(U(\vec{x}) U^\dagger(\vec{y})\right)\rangle$ of the lightlike Wilson loop in the fundamental representation as measuring a momentum kick or relative
color rotation,  more generally in QCD, the color memory effect is also captured by
higher point quadrupole, sextupole, $\cdots$ correlators as well as in both the fundamental and adjoint Wilson line representations~\cite{JalilianMarian:2004da,Blaizot:2004wv,Baier:2005dv}. Such structures are ubiquitous in high-energy QCD and can, in principle, be extracted from a variety of measurements in both DIS and hadron-hadron collisions~\cite{Dominguez:2011wm}. Interestingly, as observed in \cite{Dominguez:2011wm}, dijet measurements in both DIS off nuclei and in proton-nucleus collisions are sensitive to the Weizs\"{a}cker-Williams gluon distribution we mentioned previously. 

What's the evidence for the color dipole/color memory effect?  
 The high energy asymptotics of the Regge limit is accessible at colliders where $x <  10^{-3}$ for $Q^2\geq 1$ GeV$^2$ is routinely achieved. It is observed that CGC-based models provide a good description of small-$x$ ($x\leq 0.01$) data in DIS inclusive and exclusive electron-proton scattering measurements at HERA~\cite{Rezaeian:2012ji}. This is however not conclusive because the values of the saturation scale $Q_s$ in the proton extracted from the experiments are not much larger than non-perturbative scales where the effects of confinement may be important.  Larger values of the saturation scale may be reached by instead considering heavy ion experiments.  Namely,
because the coherence length of the dipole~\cite{Ioffe:1969kf} $l_c \gg 2R$ for $x\ll A^{-1/3}$ for a nucleus with atomic number $A$, the saturation scale receives a significant 
nuclear ``oomph'': $Q_{s}^2(A)\sim A^{1/3}$~\cite{McLerran:1993ni,Kowalski:2007rw}. Experiments of this nature were performed at RHIC, where for example deuteron-gold collisions (at $\sqrt{s}=200$ GeV/nucleon) measured correlations amongst hadron pairs flying forward in the fragmentation region of the deuteron. A CGC computation predicted~\cite{Marquet:2007vb} that the likelihood that these pairs would be azimuthally correlated back-to-back will be diminished relative to those in proton-proton collisions due to the larger color memory effect carried by the pairs. These correlations can be understood in terms of the momentum broadening phenomenon
discussed above, where the larger color memory effect characterized by the larger saturation scale in deuteron-gold collisions in comparison to proton-proton collisions leads to weaker correlations among the produced hadrons.
While the predicted effect has been observed~\cite{Adare:2011sc} and is consistent with detailed CGC computations~\cite{Lappi:2012nh}, more data varying the collision energy, the nuclear size and the transverse momenta of the produced pairs are necessary to confirm whether their systematics is consistent with the CGC. 

There have also been studies of particle correlations in proton-nucleus collisions at the LHC at the much higher center-of-mass energies of $\sqrt{s}=5.02, 8.16$ TeV/nucleon. A prediction of the CGC (first made for proton-proton collisions~\cite{Dumitru,Dumitru:2010iy}) is that of a ridgelike structure in hadron pair correlation
 functions, that is long range in the relative rapidity of the pairs but collimated in their relative azimuthal angle. These were first observed in high-multiplicity triggered proton-proton collisions at $\sqrt{s}=7$ TeV by the CMS experiment~\cite{Khachatryan:2010gv}, and a significant enhancement of the signal was subsequently seen in proton-nucleus collisions~\cite{CMS:2012qk,Chatrchyan:2013nka}. While the data for pairs with $p_T \geq Q_s$ is in agreement with CGC computations~\cite{Dusling:2013qoz}, there are features of the data that seem to indicate that ridge correlations may potentially also be due to final state scattering amongst the many produced secondaries. 
The situation is however fluid; a fairly recent review of ongoing work in theory and experiment can be found in \cite{Dusling:2015gta}. 

A likely definitive search for color memory effects is feasible in DIS of electrons off large nuclei at high energies. In this case, one has the large oomph factor of the color flux from the nucleus but can also control for final state effects by varying the squared momentum resolution $Q^2$ of the lepton probe. The prospects for extraction of the color dipole cross-section at a future Electron-Ion Collider are discussed in \cite{Aschenauer:2017jsk}. In particular, the inclusive diffractive cross-section and exclusive vector meson cross-sections are proportional to $\sigma_{\rm dipole}^2$. These display very strong systematic power law variations with $Q^2$ and the nuclear size $A$~\cite{Mantysaari:2017slo} and are therefore promising signatures for definitive discovery of the color memory effect in QCD. 

\section*{Acknowledgements}
This work is supported in part by DOE grant DE-SC0007870.
RV's work is supported by the U.S. Department of Energy, Office of Science, Office of Nuclear Physics, under Contracts No. DE-SC0012704.

\begin{appendix}
\section{Classical colored particles}

In Cartesian coordinates, the action for a classical colored particle (quark) of mass $m = 1$ charged under a non-abelian gauge field is
\begin{equation}
S = \int d\tau\left[\frac{1}{2}u^{\mu}u_{\mu}- g_{\rm YM} Q^a u^{\mu}A_{\mu}^a \right] - \frac{1}{4} \int d^4y F_{\mu\nu}^a  F^{a \mu\nu} ,
\end{equation} 
where $u^{\mu} = \frac{dx^{\mu}}{d\tau}$ is the tangent to the trajectory parametrized by $\tau$.
The equations of motion are then 
\begin{equation}
\label{u}
\frac{du_{\mu}}{d\tau} + g_{\rm YM}Q^a F_{\mu\nu}^a u^{\nu} = 0,
\end{equation}
\begin{equation}
\label{A}
\mathcal{D}_{\mu}F^{a \mu\nu}(y) =  g_{\rm YM} \int d\tau Q^a u^{\nu}\delta(x(\tau) - y)
\end{equation}
and are also known as the Wong equations \cite{Balachandran:2017jha}. Note that \eqref{A} implies that
\begin{equation}
\mathcal{D}_{\nu}\int d\tau Q u^{\nu}\delta(x(\tau) - y) = 0, \qquad \mathcal{D}_{\nu} \equiv \p_{\nu} - i g_{\rm YM} [A_{\nu},\cdot],
\end{equation}
which after integration by parts and using $\frac{d}{d\tau} = u^{\nu}\p_{\nu}$ amounts to
\begin{equation}
\label{pte}
u^{\nu}\mathcal{D}_{\nu} Q = 0.
\end{equation}
This means that the color charge $Q^a$ is parallel transported along the trajectory of the quark. For $A_{\mu}$ and $Q$ in an arbitrary representation of the colored particles, \eqref{pte} can be equivalently written as
\begin{equation}
\frac{d Q^a}{d\tau} - i g_{\rm YM} u^{\nu}[A_{\nu}, Q]^a = 0.
\end{equation}
This equation can be solved iteratively   and it follows that $Q$ evolves according to
\begin{equation}
Q \rightarrow \mathcal{U} Q \mathcal{U}^{\dagger}, \qquad \mathcal{U} = \mathcal{P} e^{i g_{\rm YM} \int d x^{\mu}A_{\mu}}.
\end{equation}
Assuming that classically, the charges $Q$ are direct products of quark color vectors
\begin{equation}
\label{tpd}
Q^a T^a_{ij} = q_i \bar{q}_j, \qquad i,j = 1,2,3, 
\end{equation}
we deduce that 
\begin{equation}
q \rightarrow \mathcal{U} q, \qquad \bar{q} \rightarrow \mathcal{U}^* \bar{q}.
\end{equation}

Note that this implies that the colors of quarks transforming in the fundamental representation evolve according to
\begin{equation}
\frac{d q}{d\tau} - i g_{\rm YM}u^{\nu} A_{\nu} q = 0.
\end{equation} 
It is also interesting to note that the form of the small-$x$ effective action in the Color Glass Condensate effective theory and its connections to Reggeon Field theory~\cite{Caron-Huot:2013fea,Bondarenko:2017ern,Hentschinski:2018rrf}  can be motivated by Wong's equations~\cite{JalilianMarian:2000ad}. 
\end{appendix}

\providecommand{\href}[2]{#2}\begingroup\raggedright\endgroup

\end{document}